%% file: main.tex
\documentclass[twocolumn]{autart}

\usepackage{graphicx}
\usepackage{subcaption}
\usepackage{etoolbox}
\usepackage{scalerel}
\usepackage{picins}

\usepackage[htt]{hyphenat}
\usepackage{siunitx}
\usepackage{pifont}
\usepackage{graphicx}
\usepackage{amsmath,textcomp}
\usepackage{tikz}

\allowdisplaybreaks

\usepackage{url}
\usepackage{algpseudocode}
\usepackage{paralist}
\usepackage{color}
\usepackage{stmaryrd}
\usepackage{epsfig} 
\usepackage{bm} 
\usepackage{listings,lstautogobble}
\usepackage{mathtools}
\usepackage{mathrsfs}
\usepackage{amsfonts}
\usepackage{newpxtext,newpxmath}
\usepackage{xspace}
\usepackage{verbatim}
\usepackage{epstopdf}
\usepackage{listings}
\usepackage{parcolumns}
\usepackage{color}
\usepackage{xspace}
\usepackage{verbatim}
\usepackage[utf8]{inputenc}
\usepackage{calrsfs}
\usepackage{rotating}
\usepackage{booktabs}
\usepackage{multirow}

\usepackage[compact]{titlesec}
\titlespacing{\section}{0pt}{1.4ex}{1.4ex}
\titlespacing{\subsection}{0pt}{0.8ex}{0.8ex}
\titlespacing{\subsubsection}{0pt}{0.4ex}{0.4ex}

\usepackage[T1]{fontenc}
\usepackage[utf8]{inputenc}

 \newtheorem{example}{Example}
\newtheorem{definition}{Definition}
 \newtheorem{theorem}{Theorem}
  \newtheorem{lemma}{Lemma} 
\usepackage{algorithm}
\usepackage{algpseudocode}
  \renewcommand{\footnotesize}{\fontsize{8pt}{11pt}\selectfont}

\DeclareMathAlphabet{\mathcal}{OMS}{cmsy}{m}{n}

\usepackage{hyperref}
\hypersetup{
    colorlinks=true,
    linkcolor=black,
    urlcolor=gray,
}
\usepackage{balance}

\newcommand{\ID}{id}

\newcommand{\operator}[1]{{\normalfont \texttt{#1}}}

\newcommand{\zono}[1]{\langle #1 \rangle}

\newcommand*\xor{\oplus}
\newcommand*\xnor{\odot}
\newcommand*\exor{\overline{\oplus}}

\newcommand*\eland{\overline{\land}} 

\newcommand*\xorsum[2]{\overset{#2}{\underset{#1}{\xor}}}
\newcommand*\andmul[2]{\overset{#2}{\underset{#1}{\Pi}}}

\makeatletter
\newcommand{\vast}{\bBigg@{4}}
\newcommand{\Vast}{\bBigg@{5}}
\makeatother

\makeatletter
\DeclareRobustCommand{\nand}{\mathbin{\mathpalette\n@and@or\land}}
\DeclareRobustCommand{\nor}{\mathbin{\mathpalette\n@and@or\lor}}

\DeclareRobustCommand{\enand}{\overline{\mathbin{\mathpalette\n@and@or\land}}}
\DeclareRobustCommand{\enor}{\overline{\mathbin{\mathpalette\n@and@or\lor}}}

\newcommand{\n@and@or}[2]{%
  \vphantom{#2}%
  \ooalign{$\m@th#1#2$\cr\hidewidth$\m@th#1\sim$\hidewidth\cr}%
}
\makeatother

\begin{document}

\begin{frontmatter}

\title{Polynomial Logical Zonotope: A Set Representation \\ for Reachability Analysis of Logical Systems}

\thanks[footnoteinfo]{Corresponding author Amr Alanwar.}

\author[tum,jacobs]{Amr Alanwar}\ead{alanwar@tum.de},    
\author[KTH]{Frank J. Jiang}\ead{frankji@kth.se}, and
\author[KTH]{Karl H. Johansson}\ead{kallej@kth.se}

\address[tum]{Technical University of Munich, Germany\vspace{-2mm}} 
\address[jacobs]{Constructor University, Germany\vspace{-2mm}}  

\address[KTH]{KTH Royal Institute of Technology, Sweden\vspace{-8mm}}


\input{Sections/1-abs.tex}

\end{frontmatter}

\input{Sections/2-intro.tex}
\input{Sections/3-prelim.tex}
\input{Sections/4-polylog.tex}

\input{Sections/4-polylogExact}

\input{Sections/5-reach.tex}

\input{Sections/6-log.tex}

\input{Sections/7-eval.tex}

\input{Sections/8-con.tex}

\bibliographystyle{plain}
\bibliography{ref}

\input{Sections/AuthorBio.tex} 
\end{document}

%% file: Sections/1-abs.tex
\begin{abstract}
In this paper, we introduce a set representation called polynomial logical zonotopes for performing exact and computationally efficient reachability analysis on logical systems. We prove that through this polynomial-like construction, we are able to perform all of the fundamental logical operations  (XOR, NOT, XNOR, AND, NAND, OR, NOR) between sets of points exactly in a reduced space, i.e., generator space with reduced complexity. Polynomial logical zonotopes are a generalization of logical zonotopes, which are able to represent up to $2^\gamma$ binary vectors using only $\gamma$ generators. Due to their construction, logical zonotopes are only able to support exact computations of some logical operations (XOR, NOT, XNOR), while other operations (AND, NAND, OR, NOR) result in over-approximations in the generator space. In order to perform all fundamental logical operations exactly, we formulate a generalization of logical zonotopes that is constructed by dependent generators and exponent matrices. While we are able to perform all of the logical operations exactly, this comes with a slight increase in computational complexity compared to logical zonotopes. To illustrate and showcase the computational benefits of polynomial logical zonotopes, we present the results of performing reachability analysis on two use cases: (1) safety verification of an intersection crossing protocol and (2) reachability analysis on a high-dimensional Boolean function. Moreover, to highlight the extensibility of logical zonotopes, we include an additional use case where we perform a computationally tractable exhaustive search for the key of a linear feedback shift register. 
\end{abstract}

%% file: Sections/2-intro.tex
\section{Introduction}
For several decades, logical systems have been used to model complex behaviors in numerous applications. By modeling a system as a collection of logical functions operating in a binary vector space, we can design models that consist of relatively simple dynamics but still capture a complex system's behavior at a sufficient level of abstraction.
Some popular approaches to modeling logical systems are finite automatons, Petri nets, and Boolean Networks (BNs). Notably, logical systems have been used to successfully model the behavior of physical systems such as gene regulatory networks~\cite{shmulevich2002probabilistic,akutsu1999identification} and robotics~\cite{roli2011design,thunberg2011boolean}. One of the most common types of logical systems, discrete-event systems, are used for analysis in a variety of applications such as communication systems~\cite{cassandras2008introduction}, manufacturing systems~\cite{schuh2015experimental}, and transportation systems~\cite{dallal2017supervisory,giua2008modeling}. 

\subsection{Motivation} 

Although logical systems have been successfully used to model complex systems with relatively simple dynamics, analyzing these models can still pose computational challenges. Notably, for logical systems defined over Boolean vector spaces with $n$ bits, analysis that needs to search or propagate the logical dynamics exhaustively may have exponential computational complexities in $n$; in other words, they suffer from the so-called "curse of dimensionality".

In the control community, reachability analysis is an essential form of analysis that often suffers from the curse of dimensionality. Reachability analysis allows us to formally verify the behavior of logical systems and provide guarantees that, for example, the system will not enter into undesired states. One of the primary challenges of reachability analysis is the need to exhaustively explore the system's state space, which grows exponentially with the number of variables. To avoid exponential computational complexity, many reachability analysis algorithms leverage representations such as Binary Decision Diagram (BDD). However, due to drawbacks that we will discuss in the next section, we propose a new family of representations we call logical zonotopes.


We are inspired by the role zonotopes and polynomial zonotopes defined over real vector spaces already play in the reachability analysis of dynamical systems~\cite{girard,conf:thesisalthoff,conf:Sparsepolynomialzonotopes}. Classical zonotopes are constructed by taking the Minkowski sum of a real vector center and a combination of real vector generators. Through this construction, a set of infinite real vectors can be represented by a finite number of generators. Then, by leveraging the fact that the Minkowski sum of two classical zonotopes can be computed through the concatenation of their generators, researchers have formulated computationally efficient approaches to reachability analysis~\cite{conf:thesisalthoff}. In binary vector spaces, logical zonotopes are able to represent up to $2^\gamma$ binary vectors using only $\gamma$ generators.
In previous work, we showed that we could apply logical operations to two logical zonotopes and compute either exact or over-approximated solutions~\cite{alanwar2022logical}. In this paper, we address cases where we need only exact solutions. We will show that by adding polynomial terms into the construction of logical zonotopes, we formulate polynomial logical zonotopes that support exact computations of all fundamental logical operations (XOR, NOT, XNOR, AND, NAND, OR, NOR).

\subsection{Related Work}
As we mentioned in the previous section, many reachability analysis algorithms avoid exponential computational complexity by leveraging BDDs. Given a proper variable ordering, BDDs can evaluate Boolean functions with linear complexity in the number of variables~\cite{conf:BDDthesis}. Due to this benefit, BDDs are widely used for verifying real-world hardware systems~\cite{conf:Univboolean} and discrete event systems~\cite{conf:effreachBDD}. While BDDs play a crucial role in verification, they have well-known drawbacks, such as requiring an externally supplied variable ordering, since, in many applications, automatically finding an optimal variable ordering is an NP-complete problem~\cite{conf:npcomplete,conf:effreachBDD}. Due to these drawbacks, BDDs are difficult to apply to a general class of logical systems.

Outside of BDDs, there are also approaches to reachability analysis for logical systems modeled as BNs, or Boolean Control Networks (BCNs) for systems with control inputs, that rely on the semi-tensor product~\cite{7454743}. However, these approaches are point-wise and work with an explicit representation of sets where all set members are explicitly enumerated, leading to reachability analysis with computational complexities that are exponential in the dimension of the system's state space. Furthermore, the structure matrix used in semi-tensor product-based approaches also grows exponentially in size with respect to the number of states and inputs, making it even more challenging to apply to high-dimensional logical systems~\cite{leifeld2019overview}.

Additionally, there is also a body of work proposing new representations with similar constructions to classical zonotopes. For example, researchers propose constrained zonotopes~\cite{conf:constrainedzono}, which enable richer analysis on systems defined over real vector spaces. However, since this representation is defined over real vector spaces, it is not directly applicable to logical systems. To generalize representations with zonotope-like constructions, Combastel proposed a formulation for functional sets with typed symbols~\cite{Combastel2022}. Under Combastel's formulation, logical zonotopes are functional sets with Boolean symbols. In this work, we overview the use of polynomial logical zonotopes for the analysis of logical systems and compare their computational performance with logical zonotopes, BCNs, and BDDs in three different use cases.

\subsection{Contributions}

The main contribution of this work is the detailed introduction of polynomial logical zonotopes and their use in reachability analysis. We are able to perform all logical operations between sets of points in a reduced space, i.e., the generator space of polynomial logical zonotopes. Extending our previous work \cite{alanwar2022logical}, in this paper, we present the formulation of polynomial logical zonotopes and discuss the trade-off of using them instead of their simpler representation: logical zonotopes. Furthermore, in computations where a polynomial logical zonotope appears repeatedly, each instance is treated independently within the framework of Minkowski logical operations. This occurrence, commonly referred to in the literature as the "dependency problem," serves as a motivation for us to address this challenge. Inspired by the work in~\cite{conf:Sparsepolynomialzonotopes,conf:combastleSymbolic}, we provide an approach to resolve the dependency problem and facilitate exact reachability analysis. Explicitly, the contributions of this work are summarized by the following contributions.
\begin{enumerate}
    \item We present our formulation of polynomial logical zonotopes as a generalization of logical zonotopes.
    \item We propose performing the logical operations between sets of points over a reduced space, i.e., the generator space of polynomial logical zonotopes in exact reachability analysis.
    \item We illustrate the computational benefits of polynomial logical zonotopes and logical zonotopes in three different applications.
\end{enumerate}
To recreate our results, readers can use our publicly available logical zonotope library, which contains code for working with both logical zonotopes and polynomial logical zonotopes\footnotemark.

\footnotetext{\href{https://github.com/aalanwar/Logical-Zonotope}{https://github.com/aalanwar/Logical-Zonotope}}

\subsection{Organization}
The remainder of the paper is organized as follows. In Section~\ref{sec:prelim}, we introduce the notation and preliminary definitions we will use throughout this work. In Section~\ref{sec:poly}, we formulate polynomial logical zonotopes and detail the different operations they support. Section~\ref{sec:approach} compares polynomial logical zonotopes with logical zonotopes and discusses the trade-off between accuracy and computational complexity. In Section~\ref{sec:eval}, we evaluate the application of polynomial logical zonotopes for verifying intersection-crossing protocols, performing reachability analysis on a high-dimensional Boolean function, and the key discovery of a linear-feedback shift register (LFSR). Finally, in Section~\ref{sec:con}, we conclude the work with a discussion about the potential of both representations and future work.   

%% file: Sections/3-prelim.tex
\section{Problem Statement and Preliminaries}\label{sec:prelim}



In this section, we introduce details about the notation used throughout this work, the problem statement, and the preliminary definitions for logical systems, and reachability analysis. 

\subsection{Notation}
 The set of natural and real numbers are denoted by $\mathbb{N}$ and $\mathbb{R}$, respectively. 
 We denote the binary set $\{0,1\}$ by $\mathbb{B}$. The XOR, NOT, OR, and AND operations are denoted by $\xor,\neg,\lor$, and $\land$, respectively. Throughout the rest of the work, with a slight abuse of notation, we omit the $\land$ from $a \land b$ and write $a \, b$ instead. The NAND, NOR, and XNOR are denoted by $\nand,\nor$, and $\xnor$, respectively. Later, we use the same notation for both the classical and Minkowski logical operators, as it will be clear when the operation is taken between sets or individual vectors. Like the classical AND operator, we will also omit the Minkowski AND to simplify the presentation. The $\andmul{k=1}{p} \alpha_{k}$ denotes the ANDing of $\alpha_{k}, \forall k=1,\dots,p$. We use the notation $\exor$ for the exact XOR between sets, and similarly, for other exact logical operations, we add a bar on top of the notation. Furthermore, $\mathbb{B}^{n \times m}$ denotes the binary $n \times m$ set. Matrices are denoted by uppercase letters, e.g., $G \in \mathbb{B}^{n \times k}$, and sets by uppercase calligraphic letters, e.g., $\mathcal{Z} \subset \mathbb{B}^{n}$. Vectors and scalars are denoted by lowercase letters, e.g., $b \in \mathbb{B}^{n }$. The identity matrix of size $n \times n$ is denoted $I_n$. We denote the Kronecker product by $\otimes$. $x \in \mathbb{B}^{n}$ is an $n \times 1$ binary vector. Given a matrix $A \in \mathbb{B}^{n \times m}$, $A_{(i,\cdot)}$ represents the $i$-th row, $A_{(\cdot,j)}$ the $j$-th column, and $A_{(i,j)}$ the $j$-th entry of row $i$. Given a discrete set $\mathcal{H} \in \{ \cdot \}^n$, $|\mathcal{H}| = n$ denotes the cardinality of the set. The $\text{max}(v_1,v_2)$ of two vectors is a vector while taking the maximum value row-wise. A matrix of ones (resp. zeros) with a size of $n \times m$ is denoted by $1_{n\times m}$ (resp. $0_{n\times m}$).


\subsection{Problem Statement}
For this work, we consider a system with a logical function $f: \mathbb{B}^{n_x} \times \mathbb{B}^{n_u} \rightarrow \mathbb{B}^{n_x}$:
\begin{align}
    x(k+1) = f\big(x(k),u(k)\big)
\label{eq:feq}
\end{align}
where $x(k) \in \mathbb{B}^{n_x}$ is the state and $u(k) \in \mathbb{B}^{n_u}$ is the control input. The logical function $f$ can consist of any combination of $\xor,\neg,\lor,\nand,\nor,\xnor,$ and $\land$. We define the reachable sets of system~\eqref{eq:feq} by the following definition.

\begin{definition}\label{def:exactreachF}(\textbf{Exact Reachable Set}) \
Given a set of initial states $\mathcal{X}_0 \subset \mathbb{B}^{n_x}$ and a set of possible inputs $\,\mathcal{U}_k \subset \mathbb{B}^{n_u}$, the exact reachable set $\mathcal{R}_{N}$ of \eqref{eq:feq} after $N$ steps is
\begin{align*}
    \mathcal{R}_{N} = \big\{ &x(N) \in \mathbb{B}^{n_x} \; \big| \; \forall k \in \{0,...,N-1\}: \\
        & x(k+1) = f\big(x(k),u(k)\big), 
        \; x(0) \in \mathcal{X}_0, u(k) \in \mathcal{U}_k \big\}.
\end{align*}
\end{definition}



We aim to compute the exact forward reachable sets of systems defined by~\eqref{eq:feq} using our new set representation, polynomial logical zonotope, as a generalization of a logical zonotope. 


\subsection{Preliminaries}

We will represent sets of states and inputs for~\eqref{eq:feq} using logical zonotopes and polynomial logical zonotopes. As will be shown, polynomial logical zonotopes are constructed using a Minkowski XOR operation, which we define as follows.

\begin{definition}\label{def:semitensor}\textbf{(Minkowski XOR)} \
Given two sets $\mathcal{L}_1$ and $\mathcal{L}_2$ of binary vectors, the Minkowski XOR is defined between every two points in the two sets as
\begin{align}
      \mathcal{L}_1 \xor \mathcal{L}_2 &= \{z_1 \xor z_2| z_1\in \mathcal{L}_1, z_2 \in \mathcal{L}_2 \}. \label{eq:xor} 
\end{align}
\end{definition}
Similarly, we define the Minkowski NOT, OR, and AND operations as follows.
\begin{align}
\neg \mathcal{L}_1  	&= \{ \neg z_1| z_1\in \mathcal{L}_1 \},  \\
\mathcal{L}_1\lor  \mathcal{L}_2 	&= \{z_1 \lor z_2| z_1\in \mathcal{L}_1, z_2 \in \mathcal{L}_2 \}, \\
\mathcal{L}_1  \mathcal{L}_2 	&= \{z_1  z_2| z_1\in \mathcal{L}_1, z_2 \in \mathcal{L}_2 \}.\label{eq:and}
\end{align}  

Then, to contextualize the introduction of polynomial logical zonotopes, we introduce a definition of the previously developed logical zonotope set representation. Inspired by the classical zonotopic set representation, which is defined in real vector space \cite{conf:zono1998}, in previous work~\cite{alanwar2022logical}, we proposed logical zonotopes as a computationally efficient set representation for binary vectors. A logical zonotope is defined as follows.
\begin{definition}(\textbf{Logical Zonotope~\cite{alanwar2022logical}}) \label{df:zono}
Given a point $c \in \mathbb{B}^{n}$ and $\gamma \in \mathbb{N}$ generator vectors in a generator matrix $G{=}[ g_{1},  {\dots},$ $g_{\gamma}]$ $\in \mathbb{B}^{n \times \gamma}$, a logical zonotope is defined as
\begin{align*}
\mathcal{L} = \Big\{ x \in \mathbb{B}^n \; \Big| \; x = c \xor \xorsum{i=1}{\gamma}  g_{i} \beta_{i}, \, \beta \in \{0,1\}^\gamma \Big\} \, .
\end{align*}
We use the shorthand notation $\mathcal{L} = \zono{c,G}$ for a logical zonotope. 
\end{definition}
Now, starting in the next section, we will extend the formulation of logical zonotopes to develop polynomial logical zonotopes and discuss the implications of the new representation.

%% file: Sections/4-polylog.tex
\section{Polynomial Logical Zonotopes}\label{sec:poly}

In this section, we present the formulation of polynomial logical zonotopes. Logical zonotopes, a special case of polynomial logical zonotopes, can only support exact computations for operators XOR, NOT, and XNOR and over-approximated computations for AND, NAND, OR, and NOR in the generator space. Since in certain applications, one might need to perform an exact analysis instead of an over-approximated one, we extend the formulation of logical zonotopes with additional constructions that enable exact computations for all of the fundamental logical operations in the generator space. In the next subsections, we introduce this new construction and how to compute the fundamental logical operations exactly. 

\subsection{Set Representation}

The polynomial logical zonotope is defined as follows.

\begin{definition}(\textbf{Polynomial Logical Zonotope}) \label{df:zonopoly}
Given a point $c \in \mathbb{B}^{n}$, a dependent generator matrix $G=\begin{bmatrix} g_{1}, \dots ,g_{h}\end{bmatrix}$ $\in \mathbb{B}^{n \times h}$, identifier $id \in  \mathbb{N}^{1 \times p }$ for identifying the dependent factors $\alpha_{1},\dots,\alpha_{p}$, and an exponent matrix $E \in \mathbb{B}^{p \times h}$, a polynomial logical zonotope is defined as
\begin{align*}
\mathcal{P} = \Big\{ & x \in \mathbb{B}^n \; \Big| \; x = c \xor \xorsum{i=1}{h} \Big(\andmul{k=1}{p} \alpha_{k}^{E_{(k,i)}} \Big) g_{i} ,\,  \alpha  \in \{0,1\}^p \Big\} \, .
\end{align*}
We use the shorthand notation $\mathcal{P} = \zono{c,G,E,id}$ for a polynomial logical zonotope. 
\end{definition}

     
Interestingly, polynomial logical zonotopes can be viewed as functional sets with Boolean symbols~\cite{Combastel2022}. We give the following example to illustrate the proposed set representation and its associated binary points. 

\renewcommand{\arraystretch}{1.05}

\begin{example}
Consider the following polynomial logical zonotope
\begin{align}\label{eq:barp1}
    \bar{\mathcal{P}}_1 = \Biggl \langle
    \begin{bmatrix}
0\\1\\0
    \end{bmatrix},\begin{bmatrix}
0 & 1\\1 & 1\\1 & 1
    \end{bmatrix},\begin{bmatrix}
1 & 1\\0 & 1
    \end{bmatrix} ,\begin{bmatrix}
1  & 2
    \end{bmatrix}  
   \Biggr \rangle. 
\end{align}
This is interpreted as  the following set
\begin{align*}
      \bar{\mathcal{P}}_1 {=} \Bigg\{ \begin{bmatrix}
 0\\1\\0
\end{bmatrix} {\xor} \begin{bmatrix}
 0\\1\\1
\end{bmatrix} \alpha_1 {\xor} \begin{bmatrix}
 1\\1\\1
\end{bmatrix} \alpha_1 \alpha_2  \Bigg| \alpha_{1},\alpha_{2} \in\{0,1\}\Bigg\}.
\end{align*}
By considering all possible binary combination of  $\alpha_{1}$ and $\alpha_{2}$, we get the following set of points
\begin{align*}
\Biggl\{
     \begin{bmatrix}
       0 \\
     0 \\
     1  
\end{bmatrix},
     \begin{bmatrix}
       0 \\
     1 \\
     0  
\end{bmatrix},
\begin{bmatrix}
     1 \\
     1\\
     0
\end{bmatrix}\Biggl\}.
\end{align*}
where the identifier vector $id = \begin{bmatrix} 1 & 2\end{bmatrix}$ stores the identifier 1 for the dependent factor $\alpha_1$ and the identifier $2$ for the dependent factor $\alpha_2$.
\end{example}


Next, we borrow the operator \operator{mergeID} from \cite{conf:Sparsepolynomialzonotopes} that is necessary in order to build a common representation of exponent matrices and fully exploit the dependencies between identical dependent factors.
\begin{prop}[Merge ID {\cite[Prob 1]{conf:Sparsepolynomialzonotopes}}] Given two polynomial logical zonotopes, $\bar{\mathcal{P}}_1 = \langle c_1, G_{1}, \bar{E}_1, \bar{\ID}_1 \rangle$ and $\bar{\mathcal{P}}_2 = \langle c_2, G_2,\bar{E}_2, \bar{\ID}_2 \rangle$, \operator{mergeID} returns two adjusted polynomial logical zonotopes $\mathcal{P}_1$ and $\mathcal{P}_2$ with identical identifier vectors that are equivalent to $\mathcal{P}_1$ and $\mathcal{P}_2$, and has a complexity of $\mathcal{O}(p_1p_2)$:
	\begin{equation*}
	\begin{split}
		& \operator{mergeID}(\bar{\mathcal{P}}_1,\bar{\mathcal{P}}_2) = \big ( \underbrace{\langle c_1,G_1,  E_1, \ID \rangle}_{\mathcal{P}_1}, \underbrace{\langle c_2, G_2, E_2, \ID \rangle}_{\mathcal{P}_2} \big ) \\
		& ~~ \\
		& \text{with} ~~ \ID = \begin{bmatrix} \bar{\ID}_1 & \bar{\ID}_{2(\mathcal{H})} \end{bmatrix},~~\mathcal{H} = \left\{ i~ |~ \bar{\ID}_{2(i)} \not\in \bar{\ID_1} \right\}, \\
		& ~~~~~~~ E_1 = \begin{bmatrix} \bar{E}_1 \\ 0_{|\mathcal{H}| \times h_1} \end{bmatrix} \in \mathbb{B}^{a \times h_1}, \\
		& ~~~~~~~ E_{2(i,\cdot)} = \begin{cases} \bar{E}_{2(j,\cdot)},~ \mathrm{if} ~ \exists j~\ID_{(i)} = \bar{\ID}_{2(j)} \\ 0_{1 \times h_2}, ~\mathrm{otherwise} \end{cases} i = 1 \dots a,
	\end{split}
	\end{equation*}
where $a = |\mathcal{H}|+p_1$. 
\label{prop:mergeID}
\end{prop}
Next, we show an example on the operator \operator{mergeID}.
\begin{example}
    Consider the following polynomial logical zonotope
\begin{align*}
     \bar{\mathcal{P}}_2 = \Biggl \langle
    \begin{bmatrix}
1\\0\\0
    \end{bmatrix},\begin{bmatrix}
1& 0\\0 & 1\\1 & 0
    \end{bmatrix},\begin{bmatrix}
0 & 1\\1 & 1
    \end{bmatrix} ,\begin{bmatrix}
1  & 3
    \end{bmatrix}  
   \Biggr \rangle.
\end{align*}
This is interpreted as the following set
\begin{align*}
 \bar{\mathcal{P}}_2 {=} \Bigg\{ \begin{bmatrix}
 1\\0\\0
\end{bmatrix} {\xor} \begin{bmatrix}
 1\\0\\1
\end{bmatrix} \alpha_2 {\xor} \begin{bmatrix}
 0\\1\\0
\end{bmatrix} \alpha_1 \alpha_2  \Bigg| \alpha_{1},\alpha_{2} \in\{0,1\}\Bigg\}.
\end{align*}
where the identifier vector $id = \begin{bmatrix} 1 & 3\end{bmatrix}$ stores the identifier 1 for the dependent factor $\alpha_1$ and the identifier $3$ for the dependent factor $\alpha_2$. If we apply the operator \operator{mergeID}($\bar{\mathcal{P}}_1, \bar{\mathcal{P}}_2$) where $\bar{\mathcal{P}}_1$ is defined in \eqref{eq:barp1}, we get the following sets with common identifiers. 
\begin{align*}
    \mathcal{P}_1 &= \Biggl \langle
    \begin{bmatrix}
0\\1\\0
    \end{bmatrix},\begin{bmatrix}
0 & 1\\1 & 1\\1 & 1
    \end{bmatrix},\begin{bmatrix}
1 & 1\\0 & 1\\ 0& 0
    \end{bmatrix} ,\begin{bmatrix}
1 &2 & 3
    \end{bmatrix}  
   \Biggr \rangle \\
        \mathcal{P}_2 &= \Biggl \langle
    \begin{bmatrix}
1\\0\\0
    \end{bmatrix},\begin{bmatrix}
1& 0\\0 & 1\\1 & 0
    \end{bmatrix},\begin{bmatrix}
0 & 1\\0 & 0\\ 1& 1
    \end{bmatrix} ,\begin{bmatrix}
1 &2 & 3
    \end{bmatrix}  
   \Biggr \rangle
\end{align*}
\end{example}

We consider as well an operator \operator{uniqueID}$(p)$ for generating unique $p$ ids out of longer repeated ids. This has a complexity of $\mathcal{O}(p)$.

Next, we provide the Minkowski logical operations using polynomial logical zonotopes. 

\subsection{Minkowski Logical Operations}
We propose to perform the following Minkowski logical operations in the generator space of polynomial logical zonotopes. Polynomial logical zonotopes are closed under all logical operations. 
\subsubsection{Minkowski XOR ($\xor$):} \,
We start with the Minkowski XOR over the generator space of polynomial logical zonotope as follows.
\begin{lemma}
\label{lem:xorpoly}
Given two polynomial logical zonotopes $\mathcal{P}_1=\zono{c_1,G_1,E_1,{id}_1}$ and $\mathcal{P}_2=\zono{c_2,G_2,E_2,{id}_2}$, the Minkowski XOR is computed as: 
\begin{align}
      \mathcal{P}_1 \xor \mathcal{P}_2  &{=} \Bigg\langle c_{1} \xor c_{2}, \begin{bmatrix} G_{1} , G_{2} \end{bmatrix}, \begin{bmatrix} E_{1} & 0 \\ 0 & E_{2} \end{bmatrix},\operator{uniqueID}(p_1 + p_2) \Bigg\rangle.
     \label{eq:xorminkpoly}
\end{align}
\end{lemma}
\begin{pf}
Let us denote the right-hand side of \eqref{eq:xorminkpoly} by $\mathcal{P}_{\xor}$. We aim to prove that $\mathcal{P}_1 \xor \mathcal{P}_2 \subseteq \mathcal{P}_{\xor}$ and $\mathcal{P}_{\xor} \subseteq \mathcal{P}_1 \xor \mathcal{P}_2$. Choose any $z_1 \in \mathcal{P}_1$ and $z_2 \in \mathcal{P}_2$ 
\begin{align*}
 \exists \hat{\alpha}_{1} &: z_1 = c_1 \xor \xorsum{i=1}{h_1} \Big(\andmul{k=1}{p_1} \hat{\alpha}_{1,k}^{E_{1,(k,i)}} \Big) g_{1,i}  \, ,\\
 \exists \hat{\alpha}_{2} &: z_2 = c_2 \xor \xorsum{i=1}{h_2} \Big(\andmul{k=1}{p_2} \hat{\alpha}_{2,k}^{E_{2,(k,i)}} \Big) g_{2,i}  \, .
 \end{align*}
 Let  $\hat{\alpha}_{\xor,1:p_{\xor}} {=} \begin{bmatrix} \hat{\alpha}_{1,1:p_{1}}\,, \hat{\alpha}_{2,1:p_{2}} \end{bmatrix}$ with $p_{\xor} {=} p_{1} {+} p_{2}$. Given that XOR is an associative and commutative gate, we have the following:
\begin{align*}
z_1 \xor z_2 & = c_{\xor} \xor \xorsum{i=1}{h_\xor} \Big(\andmul{k=1}{p_\xor} \hat{\alpha}_{\xor,k}^{E_{\xor,(k,i)}} \Big) g_{\xor,i}  \, ,
\end{align*}
where $c_{\xor} = c_{1} \xor c_{2}$, $G_{\xor} = \begin{bmatrix} G_{1}\,,\, G_{2} \end{bmatrix}$ with $G_{\xor}{=}\Big[ g_{\xor,1},$  ${\dots} ,g_{\xor,q_{\xor}}\Big]$, and $E_{\xor}=\begin{bmatrix} E_{1} & 0 \\ 0 & E_{2} \end{bmatrix}$. Thus, $z_1 \xor z_2 \in \mathcal{P}_{\xor}$ and therefore $\mathcal{P}_1 \xor \mathcal{P}_2 \subseteq \mathcal{P}_{\xor}$. Conversely, let $z_{\xor} \in \mathcal{P}_{\xor}$, then 
 \begin{align*}
 \exists \hat{\alpha}_{\xor} &: z_{\xor} = c_{\xor} \xor \xorsum{i=1}{h_\xor} \Big(\andmul{k=1}{p_\xor} \hat{\alpha}_{\xor,k}^{E_{\xor,(k,i)}} \Big) g_{\xor,i}  \, .
 \end{align*}
  Partitioning $\hat{\alpha}_{\xor,1:p_{\xor}}=\begin{bmatrix}\hat{\alpha}_{1,1:p_{1}}\,,\, \hat{\alpha}_{2,1:p_{2}}\end{bmatrix}$, it follows that there exist $z_1 \in \mathcal{P}_1$ and $z_2 \in \mathcal{P}_2$ such that $z_{\xor} = z_1 \xor z_2$. Therefore, $z_{\xor} \in \mathcal{P}_1 \xor \mathcal{P}_2$ and $ \mathcal{P}_{\xor} \subseteq \mathcal{P}_1  \xor \mathcal{P}_2$.
\end{pf}  
\subsubsection{Minkowski AND:} \,
Different from logical zonotopes, Minkowski AND can be performed exactly using polynomial logical zonotopes.

\begin{lemma}
\label{lem:andpoly}
Given two polynomial logical zonotopes $\mathcal{P}_1=\zono{c_1,G_1,E_1,{id}_1}$ and $\mathcal{P}_2=\zono{c_2,G_2,E_2,{id}_2}$, the Minkowski AND is computed exactly and leads to $\mathcal{P}_{\land}=\zono{c_\land,G_\land,E_\land,{id}_{\land}}$ where: 
\begin{align}
c_{\land} =& c_{1}  c_{2}, \label{eq:andCmink}\\
G_{\land} = &\Big[ c_{1} g_{2,1},\dots, c_{1} g_{2,h_2}, c_{2} g_{1,1},\dots, c_{2} g_{1,h_1},\nonumber \\ 
&\,\,\,\, g_{1,1} g_{2,1},g_{1,1} g_{2,2},\dots, g_{1,h_1} g_{2,h_2} \Big],\\
E_{\land} =& \Bigg[  \begin{bmatrix} 0_{p_1 \times 1}  \\ E_{2,(.,1)}  \end{bmatrix}\!,\!{...},\!\begin{bmatrix} 0_{p_1 \times 1}  \\ E_{2,(.,h_2)}   \end{bmatrix},\begin{bmatrix} E_{1,(.,1)}  \\ 0_{p_2 \times 1} \end{bmatrix}\!,\!{...},\!\begin{bmatrix} E_{1,(.,h_1)}  \\ 0_{p_2 \times 1} \end{bmatrix}, \nonumber \\
&\,\,\,\,\begin{bmatrix} E_{1,(.,1)} \\  E_{2,(.,1)}  \end{bmatrix},\begin{bmatrix} E_{1,(.,1)} \\  E_{2,(.,2)}  \end{bmatrix}\!,\!{...},\!\begin{bmatrix} E_{1,(.,h_1)} \\  E_{2,(.,h_2)}  \end{bmatrix}\Bigg], \nonumber \\
{id}_{\land} = & \,\, \operator{uniqueID}(p_1 + p_2+p_1p_2).
\end{align}
\end{lemma}

\begin{pf}
We aim to prove that $\mathcal{P}_1 \mathcal{P}_2 \subseteq \mathcal{P}_{\land}$ and $\mathcal{P}_{\land} \subseteq \mathcal{P}_1 \mathcal{P}_2$. Choose $z_1 \in \mathcal{P}_1$ and $z_2 \in \mathcal{P}_2$. Then, we have
\begin{align}
 \exists \hat{\alpha}_{1} &: z_1 = c_1 \xor \xorsum{i=1}{h_1} \Big(\andmul{k=1}{p_1} \hat{\alpha}_{1,k}^{E_{1,(k,i)}} \Big) g_{1,i}  \, ,\label{eq:pzZ1} \\
 \exists \hat{\alpha}_{2} &: z_2 = c_2 \xor \xorsum{i=1}{h_2} \Big(\andmul{k=1}{p_2} \hat{\alpha}_{2,k}^{E_{2,(k,i)}} \Big) g_{2,i} \, . \label{eq:pzZ2}
\end{align}
ANDing \eqref{eq:pzZ1} and \eqref{eq:pzZ2} results in
\begin{align}
z_1z_2=  & c_{1}  c_{2} \xor 
 \Bigg(\xorsum{i=1}{h_2} \Big(\andmul{k=1}{p_2} \hat{\alpha}_{2,k}^{E_{2,(k,i)}} \Big) g_{2,i} c_1 \Bigg)\nonumber\\ 
&\xor
\Bigg(\xorsum{i=1}{h_1} \Big(\andmul{k=1}{p_1} \hat{\alpha}_{1,k}^{E_{1,(k,i)}} \Big) g_{1,i}  c_2\Bigg) \nonumber \\
&\xor \Bigg( \xorsum{i=1_1,i_2=1}{h_1,h_2} \Big(\andmul{k_1=1}{p_1} \hat{\alpha}_{1,k_1}^{E_{1,(k_1,i_1)}} \Big) g_{1,i_1}\Big(\andmul{k_2=1}{p_2} \hat{\alpha}_{2,k_2}^{E_{2,(k_2,i_2)}} \Big) g_{2,i_2}\Bigg).
 \end{align}
 Concatenating the factors in
\begin{align}
   \hat{\alpha}_{\land} &= \big[ \hat{\alpha}_{1,1:p_1},\,\hat{\alpha}_{2,1:p_2},\, \hat{\alpha}_{1,1}\hat{\alpha}_{2,1},\dots,\hat{\alpha}_{1,p_1}\hat{\alpha}_{2,p_2}\big]
\end{align}
results in having $E_{\land}$ and $G_{\land}$. Thus, $z_1 z_2 \in \mathcal{P}_{\land}$ and therefore $\mathcal{P}_1 \mathcal{P}_2 \subseteq \mathcal{P}_{\land}$. Conversely, let $z_{\land} \in \mathcal{P}_{\land}$, then 
 \begin{align*}
 \exists \hat{\alpha}_{\land} &: z_{\land} = c_{\land} \xor \xorsum{i=1}{h_{\land}} \Big(\andmul{k=1}{p_{\land}} \hat{\alpha}_{\land,k}^{E_{\land,(k,i)}} \Big) g_{\land,i}  \, .
 \end{align*}
  Partitioning $\hat{\alpha}_{\land} {=} \big[ \hat{\alpha}_{1,1:p_1},\,\hat{\alpha}_{2,1:p_2},\, \hat{\alpha}_{1,1}\hat{\alpha}_{2,1},\dots,\hat{\alpha}_{1,p_1}\hat{\alpha}_{2,p_2}\big]$, 
  it follows that there exist $z_1 \in \mathcal{P}_1$ and $z_2 \in \mathcal{P}_2$ such that $z_{\land} = z_1 z_2$. Therefore, $z_{\land} \in \mathcal{P}_1 \mathcal{P}_2$ and thus $ \mathcal{P}_{\land} \subseteq \mathcal{P}_1   \mathcal{P}_2$.
\end{pf}

\subsubsection{Minkowski NOT ($\neg$), XNOR ($\xnor$), NAND ($\nand$), OR ($\vee$), NOR ($\nor$)} 
Based on the operations presented so far, we can compute the Minkowski NOT and XNOR as follows:
\begin{align}
     \neg \mathcal{P} &= \Big\langle c \xor 1_{n \times 1} ,G,E,id \Big\rangle,
     \label{eq:negminkpoly} \\
\mathcal{P}_1 \xnor  \mathcal{P}_2 	& =  \neg( \mathcal{P}_1  \xor \mathcal{P}_2 ). \label{eq:xnorminkpoly}
\end{align}

Using Minkowski NOT and AND operations, we can compute the Minkowski NAND ($\nand$). We can also compute Minkowski OR ($\vee$), and NOR ($\nor$) exactly using the Minkowski NAND:

\begin{align}
    \mathcal{P}_1 \nand \mathcal{P}_2 &= \neg (\mathcal{P}_1 \mathcal{P}_2),
     \label{eq:nandminkpoly}\\
    \mathcal{P}_1 \lor  \mathcal{P}_2 &= ( \neg \mathcal{P}_1 ) \nand ( \neg \mathcal{P}_2), \\
    \mathcal{P}_1 \nor  \mathcal{P}_2 &= \neg (\mathcal{P}_1 \lor \mathcal{P}_2).
\end{align}

%% file: Sections/4-polylogExact.tex
\subsection{Exact Logical Operations}
If a polynomial logical zonotope occurs several times in a calculation, each occurrence is taken independently during the Minkowski logical operations. This phenomenon is known in the literature as the "dependency problem." This motivates us to introduce an id for each factor inspired by~\cite{conf:Sparsepolynomialzonotopes,conf:combastleSymbolic} to solve the dependency problem and provide an exact reachability analysis.

We present the following logical operations that take into account the dependency between variables. We start by the exact XOR $(\exor)$ over the generator space of the polynomial logical zonotopes. We execute the exact logical operations after executing the operator \operator{mergeID} on the two input polynomial logical zonotopes. 

\subsubsection{Exact XOR ($\exor$):}
The exact XOR is performed as follows. 
\begin{lemma}
\label{lem:exactxorpoly}
Given two polynomial logical zonotopes $\mathcal{P}_1=\zono{c_1,G_1,E_1,id}$ and $\mathcal{P}_2=\zono{c_2,G_2,E_2,id}$ with a common identifier vector id, the exact XOR is computed as: 
\begin{align}
      \mathcal{P}_1 \, \exor \, \mathcal{P}_2  &= \Bigg\langle c_{1} \xor c_{2}, \begin{bmatrix} G_{1} , G_{2} \end{bmatrix}, \begin{bmatrix} E_{1} ,  E_{2} \end{bmatrix}, id \Bigg\rangle.
     \label{eq:xorexactpoly}
\end{align}
\end{lemma}

\begin{pf}
The proof is the same as Minkowski XOR while utilizing the merged identifiers instead of assigning unique identifiers.  
\end{pf}
In order to highlight the importance of the exact logical operation, we present the following motivating example of the dependency problem in set-based theory. 
\begin{example}\label{ex:xordepen}
    Consider the following polynomial logical zonotope $\mathcal{P}_3 = \zono{0,1,1,1}$, which encapsulates the points 0 and 1. The Minkowski XOR of 
    $$\mathcal{P}_3 \xor \mathcal{P}_3 = \Bigg\langle 0,\begin{bmatrix} 1 , 1\end{bmatrix},\begin{bmatrix} 1 & 0\\0 & 1\end{bmatrix},\begin{bmatrix} 1 , 2\end{bmatrix} \Bigg\rangle, $$
    which, if evaluated, results in the following set of points $\{0,1\}$. On the other hand, the exact XOR results in the following
    $$\mathcal{P}_3 \,  \exor  \, \mathcal{P}_3 = \Bigg\langle 0,0,1,1 \Bigg\rangle, $$
    which, if evaluated, results in the following set $\{0\}$. This aligns with the expected results of XORing a variable with itself. The exact XOR comes with a solution for the dependency problem. 
\end{example}
Next, we consider the exact ANDing after performing the operator \operator{mergeID}.
\subsubsection{Exact AND ($\eland$):}
\begin{lemma}
\label{lem:exactandpoly}
Given two polynomial logical zonotopes $\mathcal{P}_1=\zono{c_1,G_1,E_1,id}$ and $\mathcal{P}_2=\zono{c_2,G_2,E_2,id}$ with a common identifier vector id, the exact ANDing leads to $\mathcal{P}_{\bar{\land}}=\zono{c_{\bar{\land}},G_{\bar{\land}},E_{\bar{\land}},id}$ where: 
\begin{align}
c_{\bar{\land}} =& c_{1}  c_{2}, \label{eq:andCmink}\\
G_{\bar{\land}} = &\Big[ c_{1} g_{2,1},\dots, c_{1} g_{2,h_2}, c_{2} g_{1,1},\dots, c_{2} g_{1,h_1},\nonumber \\ 
&\,\,\,\, g_{1,1} g_{2,1},g_{1,1} g_{2,2},\dots, g_{1,h_1} g_{2,h_2} \Big],\\
E_{\bar{\land}} =& \Bigg[ E_{2,(.,1)} ,{...}, E_{2,(.,h_2)}, E_{1,(.,1)} ,{...}, E_{1,(.,h_1)}, \nonumber \\
&\,\,\,\,\text{max}\big(E_{1,(.,1)},E_{2,(.,1)}\big) ,\text{max}\big(E_{1,(.,1)},E_{2,(.,2)}\big) ,{...},\nonumber\\
&\,\,\,\, \text{max} \big(E_{1,(.,h_1)},E_{2,(.,h_2)}\big)  \Big],
\end{align}
with a row-wise max. 
\end{lemma}
\begin{pf}
The proof is the same as Minkowski AND while utilizing the merged identifiers instead of assigning unique identifiers.  
\end{pf}
The Minkowski NOT is the same as the exact NOT, as there is no generator involved in the operation. Given that we are able to perform the exact AND and NOT operations, we will be able to perform exact NAND and, thus, all the exact logical operations. 

\subsection{Containment and Generators Simplification}\label{subsec:contain_reduce}

In certain scenarios, we might need to find a polynomial logical zonotope that contains at least the given binary vectors. This is especially relevant at the beginning of any set-based analysis, where one usually starts with an initial set of binary vectors from which a polynomial logical zonotope should be computed. One way to do that is as follows.
\begin{lemma}
\label{lm:enclosepoints}
Given a list $\mathcal{S}=\{s_1,\dots,s_p \}$ of $p$ binary vectors in $\mathbb{B}^{n}$, the polynomial logical zonotope $\mathcal{P}=\zono{ c,G,  E, \ID }$ with $s_{i} \in \mathcal{P}, \forall i = \{1,\dots,p\}$, is given by
\begin{align}
    c &= s_{1}, \\
    g_{i-1} &= s_{i} \xor c,\,\, \forall i = \{2,\dots,p\},\\
    G &= [g_1, \dots, g_{p-1}],\\
    E &=  I_p, \\
    id &= \operator{uniqueID}(p).
\end{align}
\end{lemma}
\begin{pf}
By considering the truth table of all values of $\alpha$, we can find that the evaluation of $\mathcal{P}$ results in $c = s_{1}$ at one point and $g_{i-1}\xor c=s_{i} \xor c \xor c=s_{i} $, $\forall i = \{2,\dots,p\}$, at other points. 
\end{pf}

\begin{algorithm}[t]
\caption{Function \texttt{simplify} to decrease the number of generators of a polynomial logical zonotope.}
\label{alg:reduce}
\textbf{Input}: A polynomial logical zonotope $\mathcal{P}=\zono{c, G, E, id}$ with large number of $h$ dependent generators and $p$ dependent factors \\
\textbf{Output}: A polynomial logical zonotope $\mathcal{P}_s=\zono{c_s, G_s, E_s, id_s}$ with $h_s \le h$ dependent generators and $p_s \le p$ dependent factors
\begin{algorithmic}[1]
\State $c_s = c$
\State $\mathcal{S} = \texttt{evaluate}(\mathcal{P})$  // list of all binary vectors in $\mathcal{P}$ \label{ln:evalfirst}
\State $G_s= G, E_s=E, id_s=id$
\For{$i = 1:h$}
    \State $\mathcal{S}_s =  \texttt{evaluate}(\mathcal{P}\, \backslash \, g_i)$ // evaluate $\mathcal{P}$ without $g_i$
    \If{\texttt{isequal}($\mathcal{S},\mathcal{S}_s$)} 
    \label{ln:isequalcont}
      \State $G_s =$ \texttt{remove}($G_s, g_i$)
      \State $(E_s, id_s) =$ \texttt{remove}$\big( \, (E_s, id_s), E_{(.,i)} \, \big)$ // remove the corresponding column from $E_s$ and the unused identifiers
    \EndIf
\EndFor
\State $\mathcal{P}_s=\zono{c_s, G_s, E_s, id_s}$   
\end{algorithmic}
\end{algorithm}

The drawback of the technique proposed in Lemma~\ref{lm:enclosepoints} is that it results in generators with a count equal to the number of points minus one. Thus, after finding a polynomial logical zonotope containing the given binary vectors, reducing the number of generators without sacrificing any unique binary vector would be helpful. The simplification can be made with a small modification of~\cite[Algorithm 1]{alanwar2022logical}, where the dependent generators are checked for possible exclusion. This is shown in Algorithm~\ref{alg:reduce} in which we simply check the generators and corresponding exponents; however, with additional computational complexity, it is possible to also check the dependent factors for exclusion. While this approach to simplifying the polynomial logical zonotopes is straightforward, it can scale exponentially in complexity with the number of generators. An important future work will be to develop a simplification approach for polynomial logical zonotopes that leverages approximations for better computational complexities.

%% file: Sections/5-reach.tex
\subsection{Reachability Analysis}

We utilize the polynomial logical zonotope to have an exact reachable region of \eqref{eq:feq}, as defined in Definition~\ref{def:exactreachF} and presented in Algorithm~\ref{alg:reach}, as follows.

\begin{theorem}
\label{thm:reachpoly}
Given a logical function $f: \mathbb{B}^{n_x} \times \mathbb{B}^{n_u} \rightarrow \mathbb{B}^{n_x}$ in \eqref{eq:feq} and starting from initial polynomial logical zonotope $\mathcal{R}_0 \subset \mathbb B^{n_x}$, where $x(0) \in \mathcal{R}_0$, with input polynomial logical zonotope $\mathcal{U}_k \subset \mathbb B^{n_u}$, then the exact reachable region computed as
\begin{align}
    \mathcal{R}_{k+1} =  f\big(\mathcal{R}_{k},\mathcal{U}_k\big)
\end{align}
in the generator space of polynomial logical zonotopes.
\end{theorem}
\begin{pf} The logical function consists in general of XOR and NOT operations and any logical operations constructed from the NAND. $\forall x(k) \in \mathcal{R}_{k}$ and $u(k) \in \mathcal{U}_k$, we are able to compute Minkowski XOR and NOT exactly using Lemma~\ref{lem:xorpoly} and \eqref{eq:negminkpoly}, Minkowski NAND using \eqref{eq:nandminkpoly}. Moreover, we are able to compute the exact XOR and AND operations by Lemma~\ref{lem:exactxorpoly} and \ref{lem:exactandpoly}, respectively, and thus the exact NAND operation based on NOT and exact AND  operations. Given that the NAND is a universal gate, we can perform all the remaining logical operations in the generator space. 
\end{pf}

In Algorithm~\ref{alg:reach}, we overview an algorithm based on Theorem~\ref{thm:reachpoly} for $N$-step reachability analysis using polynomial logical zonotopes. First, in line~\ref{ln:contain}, we use Lemma~\ref{lm:enclosepoints} to convert the initial set of points ${\mathcal S}_0$ to get an initial polynomial logical zonotope $\bar{\mathcal{R}}_{0}$ which is further simplified to ${\mathcal{R}}_{0}$ as explained in Section~\ref{subsec:contain_reduce}. We apply the same steps to get the input polynomial logical zonotope $\mathcal{U}_{k}$. Then, we iterate $N$ times to find the $N$th-step reachable set as a polynomial logical zonotope.

\begin{algorithm}[t]
\caption{Reachability analysis for $N$-steps}
\label{alg:reach}
\textbf{Input}: A logical function $f$, an initial set of points $\mathcal{S}_0$, a set of control input points $\mathcal{S}_{u,k}$, $\forall k=1,\dots,N$\\ 
\textbf{Output}: A reachable polynomial logical zonotope ${\mathcal{R}}_{N}$ at the N-th step
\begin{algorithmic}[1]
\State $\bar{\mathcal{R}}_{0} =$ \texttt{enclosePoints}($\mathcal{S}_0$) // Enclose the set of points with a polynomial logical zonotope using Lemma~\ref{lm:enclosepoints}\label{ln:contain} 
\State  ${\mathcal{R}}_{0} =$ \texttt{simplify}($\bar{\mathcal{R}}_{0}$) // Reduce the number of generators using Algorithm~\ref{alg:reduce}
\State $\bar{\mathcal{U}}_{k} =$ \texttt{enclosePoints}($\mathcal{S}_{u,k}$), $\forall k=0,\dots,N-1$ \label{ln:containU} 
\State ${\mathcal{U}}_{k} =$ \texttt{simplify}($\bar{\mathcal{U}}_{k}$),  $\forall k=0,\dots,N-1$ \label{ln:reduceU}
  \For{$k = 0:N-1$} 
  \State  ${\mathcal{R}}_{k+1} = f({\mathcal{R}}_{k}, \mathcal{U}_k)$ // Apply Minkowski and exact logical operations						
  \EndFor
\end{algorithmic}
\end{algorithm}

%% file: Sections/6-log.tex


\section{Comparison with Logical Zonotopes}\label{sec:approach}
In this section, we compare polynomial logical zonotopes with their special case, logical zonotopes. We start by reminding the reader of the application of Minkowski XOR, NOT, XNOR, AND, NAND, OR, NOR, and reachability analysis to logical zonotopes.
Then, we discuss the computational complexity of both polynomial logical zonotopes and logical zonotopes to highlight the trade-off between the full and simpler representation. 

\subsection{Minkowski Logical Operations with Logical Zonotopes}
In this section, we briefly introduce the application of each fundamental logical operation on logical zonotopes. We separate them into two groups: (1) the Minkowski operations that yield exact solutions and (2) the Minkowski operations that yield over-approximations. For proofs and more details, we refer readers to~\cite{alanwar2022logical}.

Logical zonotopes can support exact solutions for Minkowski XOR, NOT, and XNOR operations. For logical zonotopes, $\mathcal L_1 = \langle c_1, G_1 \rangle$ and $\mathcal L_2 = \langle c_2, G_2 \rangle$, the Minkowski XOR, NOT, and XNOR are computed as follows.
\begin{align}
    \mathcal{L}_1 \xor \mathcal{L}_2  &= \Big\langle c_{1} \xor c_{2}, \begin{bmatrix} G_{1} , G_{2} \end{bmatrix} \Big\rangle,
     \label{eq:xormink}\\
    \neg \mathcal{L}_1 &= \Big\langle c_1 \xor 1_{n \times 1} ,G_1 \Big\rangle,
     \label{eq:negmink}\\
    \mathcal{L}_1 \xnor  \mathcal{L}_2 	& =  \neg( \mathcal{L}_1  \xor \mathcal{L}_2 ).
\end{align}

However, due to the limitations of their construction, logical zonotopes can only support over-approximated solutions for Minkowski AND, NAND, OR, and NOR as follows.

\begin{align}
    &\mathcal{L}_1  \mathcal{L}_2  \subseteq \big\langle c_{1}  c_{2} ,G_{\land} \big\rangle ,
    \label{eq:andmink} \\
    &\begin{aligned}
        G_{\land} {=}\Big[ & c_{1} g_{2,1},\dots, c_{1} g_{2,\gamma_2}, c_{2} g_{1,1},\dots, c_{2} g_{1,\gamma_1}\!\!,\nonumber,\\ 
        & g_{1,1}  g_{2,1}, g_{1,1}  g_{2,2}, \dots, g_{1,\gamma_1}  g_{2,\gamma_2}\Big] \nonumber ,
    \end{aligned}\\
    &\mathcal{L}_1 \nand \mathcal{L}_2 = \neg (\mathcal{L}_1 \mathcal{L}_2),
     \label{eq:nandmink}\\
    &\mathcal{L}_1 \lor  \mathcal{L}_2 	= ( \neg \mathcal{L}_1 ) \nand ( \neg \mathcal{L}_2), \\
    &\mathcal{L}_1 \nor  \mathcal{L}_2 	= \neg (\mathcal{L}_1 \lor \mathcal{L}_2).
\end{align}

The term over-approximation in binary sets with $\mathcal{L}_1 \subseteq \mathcal{L}_2$ means that $\mathcal{L}_2$ contains at least all the binary vectors contained in $\mathcal{L}_1$.

\begin{table*}[tbp]
\caption{We list out the Minkowski logical operation computational complexity for polynomial logical zonotopes and logical zonotopes. Note, for logical zonotopes, the Minkowski AND, NAND, OR, and NOR computations yield over-approximated instead of exact solutions. }
\label{tab:complexity}
\centering
\normalsize
\begin{tabular}{l c c}
\toprule
 Operation & Polynomial Logical Zonotope & Logical Zonotope~\cite{alanwar2022logical}\\
\midrule
Minkowski NOT & $\mathcal O(n)$ & $\mathcal{O}(n)$ \\
Minkowski XOR, XNOR & $\mathcal O(n + p_1 + p_2)$ & $\mathcal O (n)$ \\
Minkowski AND, NAND, OR, NOR & $\mathcal O(nh_1h_2 + p_1p_2)$ & $\mathcal{O}(n h_1 h_2)$ \\
Exact XOR, XNOR & $\mathcal{O}(n + p_1p_2)$ & - \\
Exact AND, NAND, OR, NOR & $\mathcal{O}(nh_1h_2 + p_1p_2)$ & - \\
\bottomrule
\end{tabular}
\end{table*}
\subsection{Reachability Analysis with Logical Zonotopes}

While sometimes we need to compute these operations exactly, there are other cases where over-approximating the solutions to these operations is acceptable. For example, when computing the reachable sets of a logical system to check if the system will reach unsafe sets, it is still possible to formulate safety guarantees based on over-approximated reachable sets. In these cases, the simpler logical zonotopes may be a better choice than polynomial logical zonotopes, as some operations have a greater computational complexity when applied to polynomial logical zonotopes. The following theorem, which is presented and proven in our previous work~\cite{alanwar2022logical}, summarizes the reachability analysis of logical zonotopes. 

\begin{theorem}
\label{thm:reach}
Given a logical function $f: \mathbb{B}^{n_x} \times \mathbb{B}^{n_u} \rightarrow \mathbb{B}^{n_x}$ in \eqref{eq:feq} and starting from initial logical zonotope $\mathcal{R}_0 \subset 
\mathbb B^{n_x}$, where $x(0) \in \mathcal{R}_0$, with input logical zonotope $\mathcal{U}_k \subset \mathbb B^{n_u}$, then the reachable region $\hat{\mathcal{R}}_{k+1}$ over-approximates the exact reachable set $\mathcal{R}_{k+1}$. $\hat{\mathcal{R}}_{k+1}$ is computed in the generator space of logical zonotopes.
\begin{align}
    \hat{\mathcal{R}}_{k+1} =  f \big(\hat{\mathcal{R}}_{k},\mathcal{U}_k\big) \supseteq \mathcal{R}_{k+1}.
\end{align}
\end{theorem}

In the next section, we will discuss the computational trade-offs between polynomial logical and logical zonotopes in more detail.

\subsection{Computational Complexity Trade-offs}
To compare the computational complexity of the presented logical operations, let two polynomial logical zonotopes be defined as $\mathcal{P}_1=\langle c_{1}, G_{1}, E_1, id_1 \rangle$ and $\mathcal{P}_2=\langle c_{2}, G_{2}, E_2, id_2 \rangle$, where $c_1, c_2 \in \mathbb B^n$, $G_1 \in \mathbb B^{n\times h_1}$, $E_1 \in \mathbb B^{p_1\times h_1}$, $G_2 \in \mathbb B^{n\times h_2}$, and $E_2 \in \mathbb B^{p_2\times h_2}$. Then, let two logical zonotopes be defined as $\mathcal L_1 = \langle c_1, G_1 \rangle$ and $L_2 = \langle c_2, G_2\rangle$.

We start with the computational complexity of the Minkowski logical operations. The computational complexity of NOT scales linearly with the dimension of the binary space, or $\mathcal O(n)$ for both logical and polynomial logical zonotopes. For applying XOR and XNOR to logical zonotopes $\mathcal L_1$ and $\mathcal L_2$, the complexity is also linear in the dimension of the binary space. However, for applying XOR and XNOR to polynomial logical zonotopes, in addition to scaling linearly with the dimension of the binary space, the operations also scale linearly with the sum of the number of dependent factors of $\mathcal P_1$ and $\mathcal P_2$ due to the $\texttt{uniqueID}$ operation in the XOR, giving both operations a complexity of $\mathcal O(n + p_1 + p_2)$. Similarly, for the AND, NAND, OR, NOR operations, the application of the operations to $\mathcal L_1$ and $\mathcal L_2$ has a complexity dominated by the AND operation of $\mathcal O(n h_1 h_2)$, while the application of the operations to $\mathcal P_1$ and $\mathcal P_2$ has a complexity of $\mathcal O(n h_1 h_2 + p_1 p_1)$ due to the additional $\texttt{uniqueID}$ operation. For the exact XOR, XNOR, AND, NAND, OR, and NOR operations, the complexity is determined using the same arguments as the Minkowski ones, however, with the complexity of the $\texttt{mergeID}$ operation, which has a complexity of $\mathcal O(p_1 p_2)$, instead of the $\texttt{uniqueID}$ operations. The exact logical operations cannot be applied to $\mathcal L_1$ and $\mathcal L_2$ due to the absence of identifiers.

Currently, one computational challenge when using polynomial logical zonotopes is handling simplifications in cases where the number of generators grows quickly. The simplify algorithm outlined in Algorithm~\ref{alg:reduce} depends heavily on the $\texttt{evaluate}$ function, which builds a list of all the binary vectors contained within a polynomial logical zonotope. Simple implementations of $\texttt{evaluate}$ can have complexities that grow exponentially with the number of dependent factors in the polynomial logical zonotope. As we show in the following section, there are many cases and applications that do not prohibitively suffer from this computational challenge. However, for the general application of polynomial logical zonotopes, a simplification algorithm with low computational complexity is an important future work.

In summary, we can see that the increase in complexity for applying logical operators primarily stems from the management of dependent factors. As shown in Table~\ref{tab:complexity}, when computing the Minkowski logical operations XOR, XNOR, AND, NAND, OR, and NOR for polynomial logical zonotopes, there is an increase in computational complexity. Moreover, when exact XOR, XNOR, AND, NAND, OR, NOR operations are necessary, polynomial logical zonotopes need to be used. However, in cases where over-approximations for AND, NAND, OR, NOR can be tolerated, then logical zonotopes without the exponent matrices and identifiers can be used to lower computational complexity. In the next section, we illustrate some of these computational trade-offs between logical zonotope and polynomial logical zonotope-based reachability analysis and search algorithms in our case studies.


%% file: Sections/7-eval.tex
\begin{figure}
     \centering
     \begin{subfigure}[b]{0.14\textwidth}
         \centering
         \includegraphics[width=\textwidth]{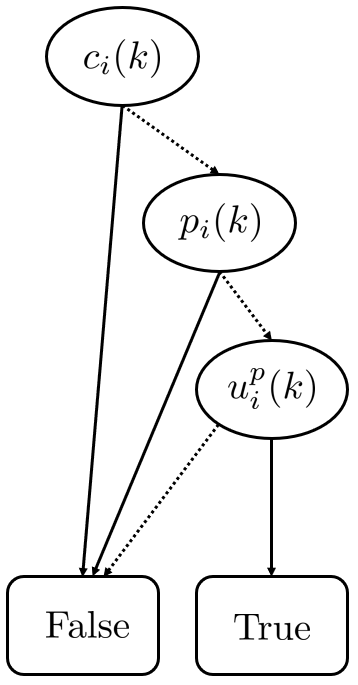}
         \caption{$p_i(k+1)$}
         \label{fig:BDD_VP}
     \end{subfigure}
     \begin{subfigure}[b]{0.165\textwidth}
         \centering
         \includegraphics[width=\textwidth]{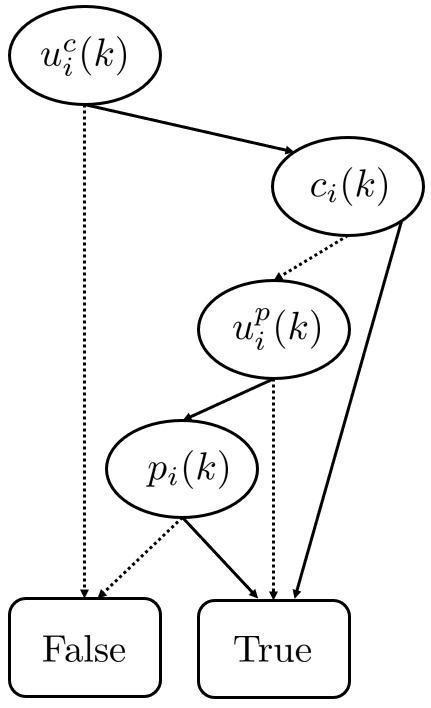}
         \caption{$c_i(k+1)$}
         \label{fig:Reduced_BDD_CF}
     \end{subfigure}
     \caption{Reduced BDDs for the intersection crossing example.}
     \label{fig:BDD}
\end{figure}

\section{Case Studies}\label{sec:eval}

\begin{table*}[tbp]
\caption{Execution Time (seconds) and number of points in each set (size) for verifying an intersection crossing protocol.}
\label{tab:exectimetrafic}
\centering
\normalsize
\begin{tabular}{c c c  c c   c c   c c}
\toprule
 &  \multicolumn{2}{c}{Zonotope} &  \multicolumn{2}{c}{Poly. Zonotope} &\multicolumn{2}{c}{BDD} & \multicolumn{2}{c}{BCN}\\
  \cmidrule(lr){2-3} \cmidrule(lr){4-5} \cmidrule(lr){6-7} \cmidrule(lr){8-9}
 Steps $N$  & Time & Size & Time & Size & Time & Size & Time & Size\\
\midrule
5 & 0.05 & 16 & 0.15 & 13 & 1.17 & 14 & 3.40 & 14 \\
10 & 0.06& 16& 0.18  &14 &3.32& 14 & 7.75&  14\\
50 & 0.15&16 & 0.25 &14&19.87 & 14& 48.40&14 \\
100 & 0.26&16 &0.45 &14&39.78 & 14&104.91 &14 \\
1000 &1.84 &16 &2.84 &14&406.60 & 14& 1142.10 &14 \\
\bottomrule
\end{tabular}
\end{table*}

To illustrate the use of operating over the generators' space of polynomial logical zonotopes and logical zonotopes, we present three different use cases. We first formulate an intersection crossing problem, where we compare the computational complexity of BDDs, BCN-based semi-tensor products, and logical zonotopes when verifying the safety of four vehicle's intersection crossing protocol. Then, we demonstrate the use of polynomial logical zonotopes for conducting reachability analysis on a high-dimensional Boolean function. In addition to the two reachability analysis use cases, we also include a use case showing how logical zonotopes can drastically improve the complexity of exhaustively searching for the key of an LFSR. All of the experiments are performed on a processor 11$^{th}$ Generation Intel(R) Core(TM) i7-1185G7 with 16.0 GB RAM. 


To compare with BCN-based semi-tensor product approaches, we use the classical definition for semi-tensor products~\cite{conf:tensorproductsurvey}. Explicitly, given two matrices $M \in \mathbb{B}^{m \times n}$ and $N  \in \mathbb{B}^{p \times q}$, the semi-tensor
product, denoted by $\ltimes$, is computed as:
\begin{align*}
    M \ltimes N = (M \otimes I_{s_1})( N \otimes I_{s_2}),
\end{align*}
with $s$ as the least common multiple of $n$ and $p$, $s_1 =s/n$, and $s_2 =s/p$. Note that we can apply a semi-tensor product to logical zonotopes as explained in~\cite{alanwar2022logical}.

\subsection{Safety Verification of an Intersection Crossing Protocol}
We extend the provided example in \cite{alanwar2022logical} to compare against polynomial logical zonotopes. More specifically, we consider an intersection where four vehicles need to pass through the intersection, while avoiding collision. For comparison, we encode their respective crossing protocols as logical functions and verify the safety of their protocols through reachability analysis using BDDs, a BCN semi-tensor product-based approach, logical zonotopes, and polynomial logical zonotopes.  
We denote whether vehicle $i$ is passing the intersection or not at time $k$ by $p_i(k)$. Then, we denote whether vehicle $i$ came first or not at time $k$ by $c_i(k)$. The control inputs $u^p_i(k)$ and $u^c_i(k)$ denote the decision of vehicle $i$ to pass or to come first at time $k$, respectively. 
For each vehicle $i=1,\dots,4$, the intersection passing protocol is represented as follows. 
\begin{align}
    {p}_i(k+1) &= u^p_i(k)  \neg {p}_i(k)  \neg c_{i}(k).
    \label{eq:VehPassSafe} 
\end{align}
Then, the logic behind coming first for each vehicle $i=1,\dots,4$ is written as follows.
\begin{align}
    c_{i}(k+1) &= \neg p_{i}(k+1) ( u^c_{i}(k) \lor ( \neg p_{i}(k) p_{i}(k+1)) ).
    \label{eq:CameFirst}
\end{align}
To perform reachability analysis, we initialize the crossing problem with the following conditions: $p_1(0) = 1, \, p_2(0) \in \{0 , 1\}, \, p_3(0) = 0 , \, p_4(0) \in \{0 , 1\},$  $c_1(0) = 1,\, c_2(0) \in \{0 , 1\}, \, c_3(0) = 0 , \, c_4(0) \in \{0 , 1\}$.
To verify the passing protocol is always safe, under any decision made by each vehicle, we perform reachability analysis under the following uncertain control inputs: $u^p_1(k)\in \{0 , 1\},\,  u^p_2(k) =0,\,u^p_3(k)\in \{0 , 1\},\,  u^p_4(k) =0,\, u^c_1(k)\in \{0 , 1\}, \, u^c_2(k) \in \{0 , 1\},\,u^c_3(k)\in \{0 , 1\},$ and $u^c_4(k) \in \{0 , 1\},\, k=0,\dots,N-1$.
%

\begin{table*}[tbp]
\caption{Execution Time (seconds) for reachability analysis of a high-dimensional Boolean function (*estimated execution times).}
\label{tab:exectimebool}
\centering
\normalsize
\begin{tabular}{c c c c c c c}
\toprule
& \multicolumn{2}{c}{Zonotope} & \multicolumn{2}{c}{Poly. Zonotope} & \multicolumn{2}{c}{BDD}  \\
 \cmidrule(lr){2-3}  \cmidrule(lr){4-5} \cmidrule(lr){6-7} 
Steps $N$ & Time & Size & Time & Size & Time & Size \\
\midrule
2 & 0.04 & 768 & 0.05 & 211 & 0.34 & 211  \\
3 & 0.05 & 896 & 0.06 & 580 &  $1.86 \times 10^5$    & 580 \\
4 & 0.06 & 896 & 0.07 &  580   &  $2.44 \times 10^6$*     &   -  \\
5 & 0.07 & 896 & 0.56 & 580    &   $> 10^6$*     & -    \\
\bottomrule
\end{tabular}
\end{table*}

We construct BDDs for each formula and execute the reduced form of the BDDs with uncertainty which is illustrated in Fig.~\ref{fig:BDD}. 
For the semi-tensor product-based approach with BCNs, we write state $x(k) {=} (\ltimes_{i=1}^4 p_{i}(k))$ $\ltimes (\ltimes_{i=1}^4 c_{i}(k))$. We write input $u(k) = (\ltimes_{i=1}^4 u^p_i(k)) \ltimes (\ltimes_{i=1}^4 u^c_i(k))$. The structure matrix $L$, which encodes \eqref{eq:VehPassSafe}-\eqref{eq:CameFirst}, is a $2^8 \times 2^{16}$ matrix where $8$ is the number of the states and $16$ is the number of states and inputs. We perform reachability analysis for the BCN using $x(k+1) = L \ltimes u(k) \ltimes x(k)$ for all possible combinations. For reachability analysis with logical zonotopes and polynomial logical zonotopes, we represent each variable in~\eqref{eq:VehPassSafe}-\eqref{eq:CameFirst} with a logical zonotope and polynomial logical zonotopes. We first compute the initial zonotope $\hat{\mathcal{R}}_0$ using Lemma~\ref{lm:enclosepoints} which contains the initial and certain states.  
Then, using Theorem~\ref{thm:reachpoly} and Theorem~\ref{thm:reach}, we compute the next reachable sets as polynomial logical zonotopes and logical zonotopes.
 
The execution time of the three approaches is presented in seconds in Table~\ref{tab:exectimetrafic}. We note that reachability analysis using logical zonotopes provides the best run-time compared to other techniques. Moreover, as the reachability analysis's time horizon increases, its run-time with logical zonotopes increases slower than the other two methods. The polynomial logical zonotopes provide exact reachability compared to logical zonotopes at the cost of slightly worse execution.  The term $p_{i}(k+1)$ appears twice in \eqref{eq:CameFirst}, which requires the exact logical operations to take care of the dependency between the terms for all iterations. On the other hand, we did not find a way to carry the dependency from one iteration $k$ to the next one $k+1$ using BDD and BCN. This is the reason for having $13$ points for polynomial logical zonotopes in comparison to $14$ for BDD and BCN with $N=5$. Notably, the impact of this dependency issue did not manifest in the subsequent steps of our example for BDD and BCN. 


   


\subsection{Reachability Analysis on a High-Dimensional Boolean Function}

We consider the following Boolean functions with $B_i \in \mathbb{B}^{10}$ and $U_i \in \mathbb{B}^{10}$, $i =1,2,3$.  
\begin{align}
    B_1(k+1) &= U_1(k) \lor ( B_2(k) \xnor B_1(k)),\\
    B_2(k+1) &= B_2(k) \xnor ( B_1(k) \land U_2(k)),\\   
    B_3(k+1) &= B_3(k) \nand ( U_2(k) \xnor U_3(k)).
\end{align}
For our reachability analysis, we initially assign sets of two possible values to $B_1(0), B_2(0)$, and $B_3(0)$. Then, we compare the execution time of reachability analysis starting from this initial condition using BDDs and logical zonotopes. We do not compare with the semi-tensor product-based approach in this example since the size structure matrix is intractable for high-dimensional systems. For the supplied variable ordering, the reachability analysis with $N=4$ and $N=5$ using BDDs was not completed in a reasonable amount of time, so we instead used the average execution time for one iteration and multiplied that time to get the total time for the reachability analysis. The results are shown in Table~\ref{tab:exectimebool}. The logical zonotopes provide a huge over-approximation in high-dimensional systems. On the other hand, polynomial logical zonotopes provide exact reachability analysis with a low execution time. Conducting reachability analysis over a high number of steps necessitates an effective reduction function to manage the number of generators, thereby improving execution time. Developing such a function will be addressed in future work. 

It is worth emphasizing that the execution times listed for polynomial logical zonotopes and logical zonotopes in Table~\ref{tab:exectimetrafic} and Table~\ref{tab:exectimebool} do not include the time spent calculating the set size. To determine the size of each set, we convert them into discrete points by considering all possible combinations of the parameters $\alpha$ and $\beta$.

\subsection{Exhaustive Search for the Key of an LFSR}
In this use case, we revisit the case presented in~\cite{alanwar2022logical} and showcase a practical application where logical zonotopes can be leveraged to decrease computational complexity while maintaining exact results.
In particular, we use logical zonotopes to reduce the search space when looking for the key of an LFSR. 

\begin{figure}
    \centering
    \includegraphics[scale=0.19]{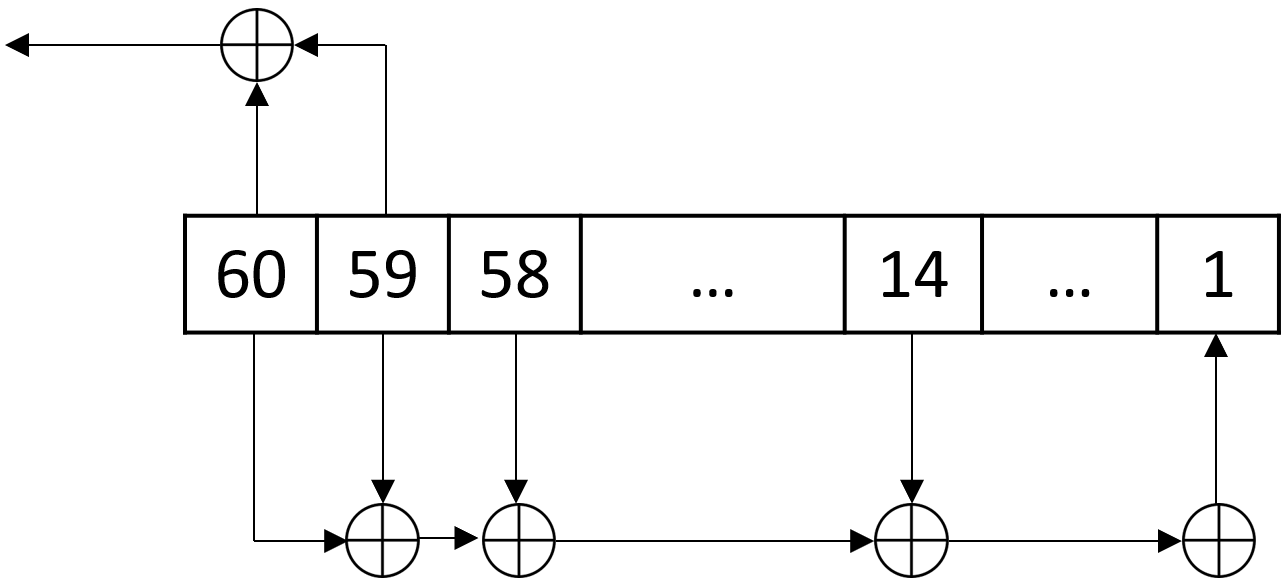}
    \caption{LFSR $A$.}
    \label{fig:LSFR}
\end{figure}

LFSRs are used intensively in many stream ciphers in order to generate pseudo-random longer keys from the input key. For simplicity we consider 60-bits LFSR $A$ initialized with the input key $K_A$ with length $l_k$. The operations on the bit level are shown in Fig.~\ref{fig:LSFR}, where
\begin{align*}
    A[1] &= A[60] \xor A[59] \xor A[58] \xor A[14], \\
    \text{output} &= A[60] \xor A[59].
\end{align*}
Each bit $i$ of the output of the LFSR is XORed with the message $m_A[i]$ to obtain one bit of the ciphertext $c_A[i]$. 

Now consider that we aim to obtain the input key $K_A$ using exhaustive search by trying out $2^{l_k}$ key values that can generate the cipher $c_A$ from $m_A$ with worst-case complexity $\mathcal{O}(2^{l_k})$ where $l_k=60$ is the key length. Instead, we propose to use logical zonotopes in Algorithm~\ref{alg:LFSR} to decrease the complexity of the search algorithm. Given that the XOR is exact already using logical zonotopes, we did not need to move forward with polynomial logical zonotopes. We start by defining a logical zonotope $\mathcal{L}_B$, which contains $0$ and $1$ in line~\ref{ln:assignlog01}. Initially, we assign a logical zonotope to each bit of LFSR $A$ in line~\ref{ln:bit2log} except the first two bits. Then, we set the first two bits of LFSR $A$ to one of the $2^2$ options of comb list in line~\ref{ln:init2bits}. Then, we call the LFSR with the assigned key bits to get a list of logical zonotopes $\mathcal{G}_A$ with misuse of notations. The pseudo-random output of logical zonotopes $\mathcal{G}_A$ is XORed with the message $m_A$ to get a list of ciphertext logical zonotopes $\mathcal{C}_A$. If any cipher of the list $c_A$ is not included in the corresponding logical zonotope $\mathcal{C}_A$, then the assigned two digits in line~\ref{ln:init2bits} are wrong, and we do not need to continue finding values for the remaining bits of LFSR $A$. We note that the \texttt{contains} function in lines \ref{ln:contains1} and \ref{ln:contains2} is implemented in the points domain by converting the logical zonotope into points and checking the containment. After finding the correct two bits with $c_A \in \mathcal{C}_A$, we continue by assigning a zero to bit by bit in line~\ref{ln:keq0}. Then we generate the pseudo-random numbers $\mathcal{G}_A$ and XORed it with the $m_A$ to get the list of cipher logical zonotopes $\mathcal{C}_A$. The ciphers' logical zonotopes $\mathcal{C}_A$ are checked to contain the list of ciphers $c_A$ and assign $\mathcal{K}_A$ in line~\ref{ln:keq1}, accordingly. 
We measured the execution time of Algorithm~\ref{alg:LFSR} with different key sizes in comparison to the execution time of traditional search in Table~\ref{tab:exectimekey}. To compute the execution time of the traditional search, we multiply the number of iterations by the average execution time of a single iteration. 


\begin{table}[tbp]
\caption{Execution Time (seconds) of exhaustive key search (*estimated execution times).}
\label{tab:exectimekey}
\centering
\normalsize
\begin{tabular}{c  c c }
\toprule
 Key Size & Algorithm~\ref{alg:LFSR} & Traditional Search \\
\midrule
30 & 1.97&  $1.18 \times 10^6$*\\
60 &4.76&   $1.26 \times 10^{15}$*\\
120 & 7.95& $1.46 \times 10^{33}$*  \\
\bottomrule
\end{tabular}
\end{table}

\begin{algorithm}[t]
\caption{Exhaustive search for LFSR key using logical zonotopes}
\label{alg:LFSR}
\textbf{Input}: A sequence of messages $m_A$ and its ciphertexts $c_A$ with length $l_m$\\ 
\textbf{Output}: The used key $\mathcal{K}_A$  with length $l_k$ in encrypting $m_A$
\begin{algorithmic}[1]
\State $\mathcal{L}_B=$\texttt{enclosePoints}$([0\,\,1])$\, // enclose the points $0$ and $1$ by a logical zonotope \label{ln:assignlog01}
\State $\text{comb}=\{00,01,10,11\}$ 
\For{$i = 3:l_k$}
 \State $\mathcal{K}_A[i]=\mathcal{L}_B$ \label{ln:keqlogZono1} // assign the logical zonotope $\mathcal{L}_B$ to the key bits \label{ln:bit2log}
  \EndFor
\For{$i = 1:4$}
  \State$\mathcal{K}_A[1:2]=\text{comb}[i]$  \label{ln:init2bits}
  \State $\mathcal{G}_A =$ \texttt{LFSR}($\mathcal{K}_A$) // generate pseudo-random numbers from the key $\mathcal{K}_A$  \label{ln:lfsr1} 
  \State $\mathcal{C}_A= \mathcal{G}_A \xor m_A$ \label{ln:enclog1}  
  \If{$\neg$\texttt{contains}\,($\mathcal{C}_A$,$c_A$)} \label{ln:contains1}
   \State \texttt{continue}; // continue if $c_A \notin \mathcal{C}_A$
  \EndIf
  \For{$j = 3:l_k$}  
  \State $\mathcal{K}_A[j]=0$. \label{ln:keq0} 
   \State $\mathcal{G}_A =$ \texttt{LFSR}($\mathcal{K}_A$) 
   \State $\mathcal{C}_A= \mathcal{G}_A \xor m_A$ \label{ln:enclog}  
  \If{$\neg$\texttt{contains}\,($\mathcal{C}_A$,$c_A$)}
  \label{ln:contains2}
   \State  $\mathcal{K}_A[i]=1$ // assign if $c_A \notin \mathcal{C}_A$ \label{ln:keq1}
  \EndIf
  \EndFor
  \If{\texttt{isequal}\,($\mathcal{K}_A \xor m_A$,$c_A$)}
  \label{ln:isequal}
   \State \texttt{return} $\mathcal{K}_A$ \label{ln:return}
  \EndIf
  \EndFor
  \end{algorithmic}
\end{algorithm}

%% file: Sections/8-con.tex
\section{Conclusion}\label{sec:con}
In this work, we propose the use of a generalization of logical zonotopes called polynomial logical zonotopes for reachability analysis on logical systems. Polynomial logical zonotopes are constructed with additional dependent generators and exponent matrices, which allow for the exact computation of the logical operations AND, NAND, OR, and NOR. In two different use cases, we show that polynomial logical zonotopes can be used for computationally efficient reachability analysis. Then, to illustrate the extensibility of logical zonotopes, we use them to reduce the computational complexity of exhaustive searches on logical systems. Moreover, we detail the trade-off between computational complexity and precision when using polynomial logical zonotopes or logical zonotopes in a computation. In future work, we will continue exploring the practical application of polynomial logical zonotopes and investigate new approaches for generator reduction. Furthermore, we will explore not only new use cases for polynomial logical zonotope-based reachability analysis but also other forms of analysis that benefit from the representation, such as search algorithms.



 \section*{Acknowledgement}

 This paper has received funding from 
 Knut and Alice Wallenberg Foundation Wallenberg Scholar Grant,
 the Swedish Research Council Distinguished Professor Grant 2017-01078, and the  Wallenberg AI, Autonomous Systems and Software Program (WASP) funded by the Knut and Alice Wallenberg Foundation.

\balance

%% file: Sections/AuthorBio.tex
\par\noindent 
\parbox[t]{\linewidth}{
\noindent\parpic{\includegraphics[height=1.5in,width=1in,clip,keepaspectratio]{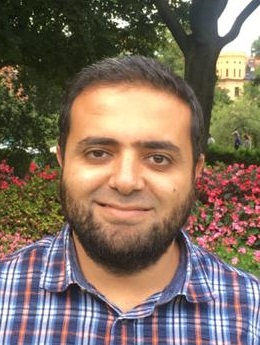}}
\noindent {\bf Amr Alanwar}\
is an assistant professor at Technical University of Munich, Germany, and an adjunct assistant professor at Constructor University, Germany. He received an M.Sc. in Computer Engineering from Ain Shams University, Cairo, Egypt, in 2013 and a Ph.D. in Computer Science from the Technical University of Munich in 2020. He was a postdoctoral researcher at KTH Royal Institute of Technology. He was also a research assistant at the University of California, Los Angeles. He received the Best Demonstration Paper Award at the 16th ACM/IEEE International Conference on Information Processing in Sensor Networks (IPSN/CPSWeek 2017) and was a finalist in the Qualcomm Innovation Fellowship for two years in a row.}

\par\noindent 
\parbox[t]{\linewidth}{
\noindent\parpic{\includegraphics[height=1.5in,width=1in,clip,keepaspectratio]{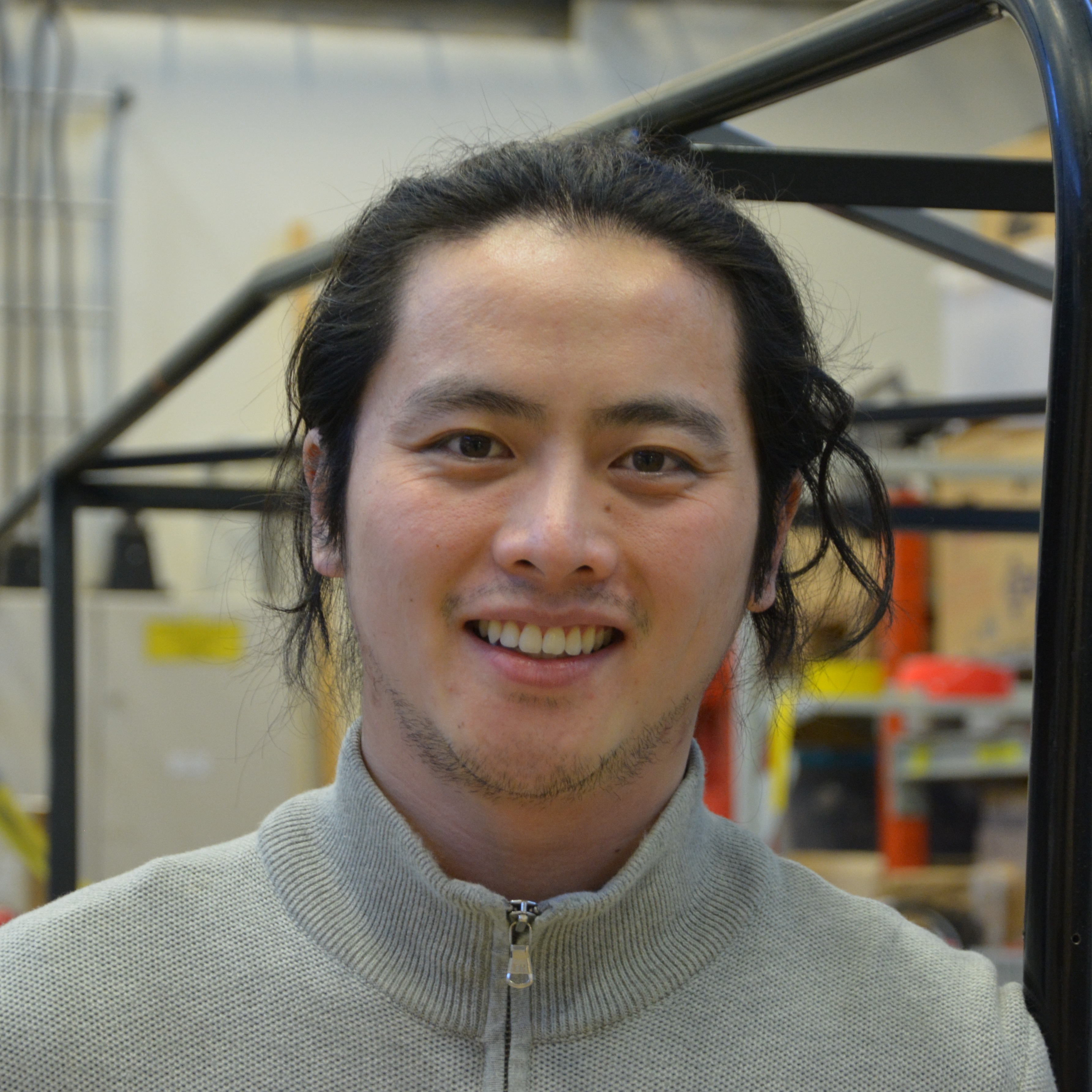}}
\noindent {\bf Frank J. Jiang}\
is a doctoral student with the School of Electrical Engineering and Computer Science at KTH Royal Institute of Technology in Sweden. He received his B.S. degree in Electrical Engineering and Computer Science from the University of California, Berkeley, in 2016 and his M.S. degree in Systems, Control, and Robotics from the KTH Royal Institute of Technology in 2019. His research interests are in formal verification, machine learning, and control, and their applications in robotics and intelligent transportation systems. He received the Best Student Paper Award at the 2020 IFAC Conference on Cyber-Physical-Human Systems (CPHS 2020).}

\par\noindent 
\parbox[t]{\linewidth}{
\noindent\parpic{\includegraphics[height=1.5in,width=1in,clip,keepaspectratio]{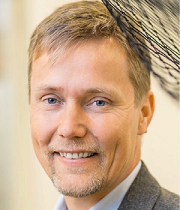}}
\noindent {\bf Karl H. Johansson}\
is the Director of Digital Futures and a Professor at the School of Electrical Engineering and Computer Science at KTH Royal Institute of Technology in Sweden. He received an M.Sc. degree in Electrical Engineering and a Ph.D. in Automatic Control from Lund University. He has held visiting positions at UC Berkeley, California Institute of Technology, Nanyang Technological University, Institute of Advanced Studies Hong Kong University of Science and Technology, Norwegian University of Science and Technology, and Zhejiang University. At KTH he directed the ACCESS Linnaeus Centre 2009-2016 and the Strategic Research Area ICT TNG 2013-2020. His research interests are in networked control systems and cyber-physical systems with applications in transportation, energy, and automation networks, areas in which he has co-authored more than 800 journal and conference papers and supervised almost 100 postdocs and Ph.D. students. He has co-authored and edited 4 books, 33 book chapters, and 7 patents. He is President of the European Control Association and a member of the IFAC Council and has served on the IEEE Control Systems Society Board of Governors and the Swedish Scientific Council for Natural Sciences and Engineering Sciences. He is a Fellow of the IEEE and the Royal Swedish Academy of Engineering Sciences.
}

%% file: main.bbl
\begin{thebibliography}{10}

\bibitem{akutsu1999identification}
Tatsuya Akutsu, Satoru Miyano, and Satoru Kuhara.
\newblock Identification of genetic networks from a small number of gene
  expression patterns under the boolean network model.
\newblock In {\em Biocomputing'99}, pages 17--28. World Scientific, 1999.

\bibitem{alanwar2022logical}
Amr Alanwar, Frank~J. Jiang, Samy Amin, and Karl~H. Johansson.
\newblock Logical zonotopes: A set representation for the formal verification
  of boolean functions.
\newblock In {\em 62nd IEEE Conference on Decision and Control}, pages 60--66,
  2023.

\bibitem{conf:thesisalthoff}
Matthias Althoff.
\newblock {\em {Reachability analysis and its application to the safety
  assessment of autonomous cars}}.
\newblock PhD thesis, Technische Universit{\"{a}}t M{\"{u}}nchen, 2010.

\bibitem{conf:Univboolean}
Jesse Bingham.
\newblock Universal boolean functional vectors.
\newblock In {\em Formal Methods in Computer-Aided Design}, pages 25--32. IEEE,
  2015.

\bibitem{conf:effreachBDD}
Martin Byrod, Bengt Lennartson, Arash Vahidi, and Knut Akesson.
\newblock Efficient reachability analysis on modular discrete-event systems
  using binary decision diagrams.
\newblock In {\em 2006 8th International Workshop on Discrete Event Systems},
  pages 288--293. IEEE, 2006.

\bibitem{conf:npcomplete}
Gianpiero Cabodi, Paolo Camurati, Luciano Lavagno, and Stefano Quer.
\newblock Disjunctive partitioning and partial iterative squaring: An effective
  approach for symbolic traversal of large circuits.
\newblock In {\em Proceedings of the 34th annual Design Automation Conference},
  pages 728--733, 1997.

\bibitem{cassandras2008introduction}
Christos~G Cassandras and St{\'e}phane Lafortune.
\newblock {\em Introduction to discrete event systems}.
\newblock Springer, 2008.

\bibitem{conf:tensorproductsurvey}
Daizhan Cheng, Hongsheng Qi, and Ancheng Xue.
\newblock A survey on semi-tensor product of matrices.
\newblock {\em Journal of Systems Science and Complexity}, 20(2):304--322,
  2007.

\bibitem{Combastel2022}
Christophe Combastel.
\newblock Functional sets with typed symbols: Mixed zonotopes and polynotopes
  for hybrid nonlinear reachability and filtering.
\newblock {\em Automatica}, 143:110457, 2022.

\bibitem{conf:combastleSymbolic}
Christophe Combastel and Ali Zolghadri.
\newblock A distributed kalman filter with symbolic zonotopes and unique
  symbols provider for robust state estimation in cps.
\newblock {\em International Journal of Control}, 93(11):2596--2612, 2020.

\bibitem{dallal2017supervisory}
Eric Dallal, Alessandro Colombo, Domitilla Del~Vecchio, and St{\'e}phane
  Lafortune.
\newblock Supervisory control for collision avoidance in vehicular networks
  using discrete event abstractions.
\newblock {\em Discrete Event Dynamic Systems}, 27(1):1--44, 2017.

\bibitem{girard}
Antoine Girard.
\newblock Reachability of uncertain linear systems using zonotopes.
\newblock In Manfred Morari and Lothar Thiele, editors, {\em Hybrid Systems:
  Computation and Control}, pages 291--305, Berlin, Heidelberg, 2005. Springer
  Berlin Heidelberg.

\bibitem{giua2008modeling}
Alessandro Giua and Carla Seatzu.
\newblock Modeling and supervisory control of railway networks using petri
  nets.
\newblock {\em Transactions on automation science and engineering},
  5(3):431--445, 2008.

\bibitem{conf:BDDthesis}
Alan~John Hu.
\newblock {\em Techniques for efficient formal verification using binary
  decision diagrams}.
\newblock stanford university, 1996.

\bibitem{conf:Sparsepolynomialzonotopes}
Niklas Kochdumper and Matthias Althoff.
\newblock Sparse polynomial zonotopes: A novel set representation for
  reachability analysis.
\newblock {\em Transactions on Automatic Control}, 66(9):4043--4058, 2020.

\bibitem{conf:zono1998}
Wolfgang K{\"{u}}hn.
\newblock Rigorously computed orbits of dynamical systems without the wrapping
  effect.
\newblock {\em Computing}, 61(1):47--67, 1998.

\bibitem{leifeld2019overview}
Thomas Leifeld, Zhihua Zhang, and Ping Zhang.
\newblock Overview and comparison of approaches towards an algebraic
  description of discrete event systems.
\newblock {\em Annual Reviews in Control}, 48:80--88, 2019.

\bibitem{7454743}
Fangfei Li and Yang Tang.
\newblock Robust reachability of boolean control networks.
\newblock {\em IEEE/ACM Transactions on Computational Biology and
  Bioinformatics}, 14(3):740--745, 2017.

\bibitem{roli2011design}
Andrea Roli, Mattia Manfroni, Carlo Pinciroli, and Mauro Birattari.
\newblock On the design of boolean network robots.
\newblock In {\em European Conference on the Applications of Evolutionary
  Computation}, pages 43--52, 2011.

\bibitem{schuh2015experimental}
Melanie Schuh, Markus Zgorzelski, and Jan Lunze.
\newblock Experimental evaluation of an active fault--tolerant control method.
\newblock {\em Control Engineering Practice}, 43:1--11, 2015.

\bibitem{conf:constrainedzono}
Joseph~K Scott, Davide~M Raimondo, Giuseppe~Roberto Marseglia, and Richard~D
  Braatz.
\newblock Constrained zonotopes: A new tool for set-based estimation and fault
  detection.
\newblock {\em Automatica}, 69:126--136, 2016.

\bibitem{shmulevich2002probabilistic}
Ilya Shmulevich, Edward~R Dougherty, Seungchan Kim, and Wei Zhang.
\newblock Probabilistic boolean networks: a rule-based uncertainty model for
  gene regulatory networks.
\newblock {\em Bioinformatics}, 18(2):261--274, 2002.

\bibitem{thunberg2011boolean}
Johan Thunberg, Petter {\"O}gren, and Xiaoming Hu.
\newblock A boolean control network approach to pursuit evasion problems in
  polygonal environments.
\newblock In {\em International Conference on Robotics and Automation}, pages
  4506--4511. IEEE, 2011.

\end{thebibliography}
